\documentclass[twocolumn,superscriptaddress]{revtex4-2}

\usepackage{graphicx}
\usepackage{dcolumn}
\usepackage{bm}

\usepackage{color}
\usepackage{amsmath}
\usepackage{float}
\usepackage{booktabs}
\usepackage{cancel}
\usepackage{caption}
\usepackage{ragged2e}  
\DeclareUnicodeCharacter{2550}{\ensuremath{\equiv}}
\usepackage[pdftex]{hyperref}
\definecolor{darkred}{rgb}{0.25,0,0}
\definecolor{darkgreen}{rgb}{0,0.25,0}
\definecolor{darkblue}{rgb}{0,0,1}
\hypersetup{colorlinks,linkcolor=darkblue,filecolor=black,urlcolor=darkblue,citecolor=darkblue}
\usepackage{multirow}
\usepackage{textcomp}
\usepackage[compact]{titlesec}
\usepackage[normalem]{ulem}
\vspace{-0.5cm} 
\DeclareUnicodeCharacter{2212}{-}
\DeclareUnicodeCharacter{0393}{-}
\begin{document}

\preprint{APS/123-QED}

\title{Over One Order of Magnitude Enhancement in Hole Mobility of 2D III-V Semiconductors through Valence Band Edge Shift}


\author{Jianshi Sun}%
\affiliation{Institute of Micro/Nano Electromechanical System and Integrated Circuit, College of Mechanical Engineering, State Key Laboratory of Advanced Fiber Materials, Donghua University, Shanghai 201620, China}

\author{Shouhang Li}
\email{shouhang.li@universite-paris-saclay.fr}
\affiliation{Centre de Nanosciences et de Nanotechnologies, CNRS, Université Paris-Saclay, 10 Boulevard Thomas Gobert, Palaiseau 91120, France}
\collaboration{These authors contributed equally to this work.}

\author{Cheng Shao}
\affiliation{Thermal Science Research Center, Shandong Institute of Advanced Technology, Jinan, Shandong 250103, China}

\author{Zhen Tong}
\affiliation{School of Advanced Energy, Sun Yat-Sen University, Shenzhen 518107, China}

\author{Meng An}
\affiliation{Department of Mechanical Engineering, The University of Tokyo, 7-3-1 Hongo, Bunkyo, Tokyo 113-8656, Japan}

\author{Yue Hu}
\affiliation{CTG Wuhan Science and Technology Innovation Park, China Three Gorges Corporation, Wuhan 430010, China}

\author{Xiongfei Zhu}
\affiliation{Institute of Micro/Nano Electromechanical System and Integrated Circuit, College of Mechanical Engineering, State Key Laboratory of Advanced Fiber Materials, Donghua University, Shanghai 201620, China}

\author{Thomas Frauenheim}
\affiliation{School of Science, Constructor University, Bremen 28759, Germany}
\affiliation{Institute for Advanced Study, Chengdu University, Chengdu 610106, China}

\author{Xiangjun Liu}%
\email{xjliu@dhu.edu.cn}
\affiliation{Institute of Micro/Nano Electromechanical System and Integrated Circuit, College of Mechanical Engineering, State Key Laboratory of Advanced Fiber Materials, Donghua University, Shanghai 201620, China}

\date{\today}%

\begin{abstract}
Two-dimensional (2D) semiconductors show great potential to sustain Moore's law in the era of ultra-scaled electronics. However, their scalable applications are severely constrained by low hole mobility. In this work, we take 2D-GaAs as a prototype of III-V semiconductors to investigate the effects of quantum anharmonicity (QA) on hole transport, employing the stochastic self-consistent harmonic approximation assisted by the machine learning potential. It is found that the room-temperature hole mobility of 2D-GaAs is reduced by $\sim$44\% as the QA effects are incorporated, which is attributed to the enhanced electron-phonon scattering from the out-of-plane acoustic polarization. The valence band edge shift (VBES) strategy is proposed to increase the hole mobility by $\sim$1600\% at room temperature, which can be realized by 1\% biaxial compressive strain. The electron-phonon scattering rate is dramatically decreased due to the full filtering of the original interband electron-phonon scattering channels that existed in the flat hole pocket. The VBES strategy can be further extended to other 2D III-V semiconductors to promote their hole mobilities.
\end{abstract}

\maketitle

Two-dimensional (2D) semiconductors have attracted considerable interest since they are expected to be ideal materials for transistors in the era of ultra-scaled post-silicon technologies\cite{waltl2022perspective,fiori2014electronics}. High carrier mobility ($\mu$) is a critical benchmark for developing higher-performance and lower-power-consumption logic devices and radio-frequency applications\cite{irds2023}. However, in most 2D semiconductors, hole mobilities ($\mu_h$) are an order of magnitude lower than electron mobilities ($\mu_e$)\cite{ponce2023long,ponce2023accurate}. This phenomenon is particularly pronounced in 2D III–V semiconductors\cite{zhang2023two}. Taking 2D gallium arsenide (GaAs) as an example, previous theoretical prediction showed that its room-temperature $\mu_e$ can be as high as 813 $\mathrm{cm\textsuperscript{2}/(Vs)}$, while its $\mu_h$ is only 14 $\mathrm{cm\textsuperscript{2}/(Vs)}$\cite{zhou2024guidelines}. Therefore, it is essential to enhance the $\mu_h$ of 2D III-V semiconductors for their scalable applications.

Strain engineering has been shown to be effective in increasing both the $\mu_e$ and $\mu_h$ of silicon\cite{thompson2006uniaxial}, which, to a large extent, enables the continuation of Moore’s law\cite{lundstrom2003moore}. Recent experimental and theoretical studies demonstrate this strategy also greatly promotes the $\mu_h$ of wurtzite GaN\cite{ponce2019route}, 4H-SiC\cite{sun2024giant}, diamond\cite{sun2025unlocking}, and rocksalt ScN\cite{rudra2023reversal}. The enhancement of $\mu_h$ in those 3D semiconductors can all be attributed to the shift of valence bands near the valence band maximum (VBM) and suppresses the interband electron-phonon scattering. Strain engineering is particularly well-suited for 2D materials, given their exceptional elasticity, high Young’s modulus, and remarkable sensitivity to external perturbations\cite{dai2019strain}. However, this strategy is below expectations when it is applied to 2D compound semiconductors \cite{cheng2020two}. In addtition, current researches primarily focus on enhancing the $\mu_e$, further amplifying the gap between $\mu_e$ and $\mu_h$ in 2D semiconductors\cite{datye2022strain,ge2014effect,balaghi2021high,sohier2019valley,yang2024biaxial}. Studies on the modulation of $\mu_h$ are limited, and it is still unclear whether strain engineering can efficiently enhance the $\mu_h$ of 2D semiconductors.

It has been possible to predict the phonon-limited electron transport properties using mode-level first-principles calculations\cite{liu2017first,yao2023intrinsic,deng2023phonon,zhou2019predicting} due to the advancements in modern algorithms and computational resources. The harmonic dynamical matrices are adopted in most available toolkits, like EPW\cite{ponce2016epw}, EPIC STAR\cite{deng2020epic}, and Phoebe\cite{cepellotti2022phoebe}, to predict the $\mu_e$ and $\mu_h$. However, the harmonic approximation fails in ferroelectrics\cite{ranalli2023temperature,verdi2023quantum,zhou2018electron,ranalli2024electron}, charge-density wave\cite{bianco2020weak}, and hydrogen-rich\cite{belli2025efficient} systems due to the presence of imaginary frequencies associated with soft phonon modes. Therefore, it is natural to incorporate quantum anharmonicity (QA) effects to eliminate imaginary frequencies and obtain accurate phonon dispersions for those systems. However, there are no imaginary frequencies in the phonon dispersions of most 2D semiconductors from the harmonic approximation. The QA effects are tacitly ignored in previous studies for 2D semiconductors \cite{cheng2020two,zhou2024guidelines,zhang2023two}. Despite its potential importance, the influence of QA effects on charge transport in 2D semiconductors has not yet been established.

In this Letter, we take 2D-GaAs as a prototype and show that the QA effects strongly renormalize the phonon spectrum, particularly the out-of-plane acoustic (ZA) polarization, which governs hole mobility. The temperature-dependent dynamical matrices are obtained through the stochastic self-consistent harmonic approximation (SSCHA) assisted by the machine learning potential. The first-principles calculations show that $\mu_h$ of 2D-GaAs is greatly reduced by $\sim$44\% as the QA effects are incorporated into the dynamical matrices. We propose that the $\mu_h$ can be increased by $\sim$16 times using the 1\% biaxial compressive strain engineering and bringing it to the same magnitude as $\mu_e$. Unlike the mechanism in 3D semiconductors mentioned above, the dramatic enhancement is mainly attributed to the valence band edge shift, which simultaneously increases both the electron relaxation time and group velocity. In addition, it is found that the ZA phonon scattering is severely suppressed for 2D-GaAs under compressive strains, and the high-frequency optical polarizations become the dominant scattering sources.

\textit{First principles theory} - According to the linearized electron Boltzmann transport equation, the $\mu$ can be expressed as\cite{ziman2001electrons},
\begin{equation}
    \mu_{\alpha \beta}=\frac{e}{n_{c} \Omega} \sum_{n} \int \frac{d^{3} \mathbf{k}}{\Omega_{\mathrm{BZ}}} v_{n \mathbf{k}, \alpha} \partial_{E_{\beta}} f_{n \mathbf{k}},
    \label{SE1}
\end{equation}
where $\alpha$ and $\beta$ are Cartesian coordinates, $\textit{e}$ is the elementary charge, and $n_{\textit{c}}$ is the charge carrier concentration determined by the Fermi level\cite{li2019resolving}. $\Omega$ and $\Omega_{\mathrm{BZ}}$ denote the volumes of the primitive cell and first Brillouin zone, respectively. $v_{n \mathbf{k}, \alpha}=\hbar^{-1} \partial \varepsilon_{n \mathbf{k}} / \partial k_{\alpha}$ is the electron group velocity of the electron mode corresponding to the Kohn-Sham energy $\varepsilon_{n \mathbf{k}}$, band index $n$, and wavevector $\mathbf{k}$. $\partial_{E_{\beta}} f_{n \mathbf{k}}$ is the perturbation to the Fermi-Dirac distribution due to the external electric field $\mathbf{E}$, and it is described as,
\begin{equation}
\begin{aligned}
\partial_{E_{\beta}} f_{n \mathbf{k}} = & \, e \frac{\partial f_{n \mathbf{k}}^{0}}{\partial \varepsilon_{n \mathbf{k}}} v_{n \mathbf{k}, \beta} \tau_{n \mathbf{k}} \\
& + \frac{2 \pi \tau_{n \mathbf{k}}}{\hbar} \sum_{m \nu} \int \frac{d \mathbf{q}}{\Omega_{\mathrm{BZ}}} \left| g_{m n \nu}(\mathbf{k}, \mathbf{q}) \right|^{2} \\
& \times \left[ \left(n_{\mathbf{q} \nu}^{0} + 1 - f_{n \mathbf{k}}^{0}\right) \delta\left(\Delta \varepsilon_{\mathbf{k}, \mathbf{q}}^{n m} + \hbar \omega_{\mathbf{q} \nu}\right) \right. \\
& \quad + \left. \left(n_{\mathbf{q} \nu}^{0} + f_{n \mathbf{k}}^{0}\right) \delta\left(\Delta \varepsilon_{\mathbf{k}, \mathbf{q}}^{n m} - \hbar \omega_{\mathbf{q} \nu}\right)\right] \partial_{E_{\beta}} f_{m \mathbf{k} + \mathbf{q}},
\end{aligned}
\label{SE2}
\end{equation}
where $\Delta \varepsilon_{\mathbf{k}, \mathbf{q}}^{n m}=\varepsilon_{n \mathbf{k}}-\varepsilon_{m \mathbf{k}+\mathbf{q}}$, $f_{n \mathbf{k}}^{0}$ is the Fermi-Dirac distribution, $n_{\mathbf{q} \nu}^{0}$ is the Bose-Einstein distribution, and $\omega_{\mathbf{q}\nu}$ is the phonon frequency.

The electron−phonon matrix elements $g_{m n \nu}(\mathbf{k}, \mathbf{q})=\left(\hbar / 2 \omega_{\mathbf{q} \nu}\right)^{1 / 2}\left\langle m \mathbf{k}+\mathbf{q}\left|\Delta_{\mathbf{q} \nu} V\right| n \mathbf{k}\right\rangle$, which quantify the probability amplitude for scattering between the electronic states $n \mathbf{k}$ and $m \mathbf{k}+\mathbf{q}$, with $\Delta_{\mathbf{q} \nu} V$ the first-order differential of the Kohn−Sham potential with respect to atomic displacement. In traditional calculations, the phonon energies and eigenvectors are obtained by diagonalizing the dynamical matrices from the harmonic approximation. We account for the QA effects and renormalize the harmonic dynamical matrices based on the SSCHA calculations\cite{monacelli2021stochastic}. The electron-phonon scattering rate (the inverse of phonon-limited electron relaxation time $\tau_{n \mathbf{k}}$) is computed within the Fan-Migdal self-energy relaxation time approximation (SERTA),\cite{giustino2017electron}
\begin{equation}
\begin{aligned}
\frac{1}{\tau_{n \mathbf{k}}} = & \frac{2 \pi}{\hbar} \sum_{m \mathbf{k}+\mathbf{q}} \left| g_{m n \nu}(\mathbf{k}, \mathbf{q}) \right|^{2} \times \\
& \left\{ \left[n_{\mathbf{q} \nu}^{0} + f_{m \mathbf{k}+\mathbf{q}}^{0}\right] \delta\left(\Delta \varepsilon_{\mathbf{k}, \mathbf{q}}^{n m} + \hbar \omega_{\mathbf{q} \nu}\right) \right. \\
& \quad + \left. \left[n_{\mathbf{q} \nu}^{0} + 1 - f_{m \mathbf{k}+\mathbf{q}}^{0}\right] \delta\left(\Delta \varepsilon_{\mathbf{k}, \mathbf{q}}^{n m} - \hbar \omega_{\mathbf{q} \nu}\right) \right\}.
\end{aligned}
\label{SE3}
\end{equation}
The Dirac delta functions represent energy conservation during the scattering processes, and the adaptive broadening approach is employed for their calculation\cite{macheda2018magnetotransport}. The SERTA result from Eq. \ref{SE2} is adopted as the initial value of $\partial_{E_{\beta}} f_{n \mathbf{k}}$, and then we iteratively solve Eqs. \ref{SE1} and \ref{SE2}, ultimately obtaining the converged value of $\mu_h$.

The Quantum Espresso package\cite{giannozzi2009quantum} is employed for the density functional theory (DFT) calculations with the Perdew−Burke−Ernzerhof (PBE) form of the exchange−correlation functional\cite{perdew2008restoring} and optimized full relativistic norm-conserving pseudopotentials\cite{hamann2013optimized} from PseudoDojo\cite{van2018pseudodojo}. The electronic band structures calculated using the PBE form of the exchange-correlation functional with full relativistic Optimized Norm-Conserving Vanderbilt (ONCV)\cite{hamann2013optimized}, Projector-Augmented Wave (PAW)\cite{blochl1994projector}, ultrasoft (US) \cite{kresse1999ultrasoft} pseudopotentials, as well as the screened Heyd-Scuseria-Ernzerhof (HSE) hybrid functional\cite{heyd2003hybrid} are shown in Section I of the Supplemental Material. The variations in the valence band edges are marginal. Therefore, we adopt the PBE form of the exchange-correlation functional and full relativistic ONCV pseudopotentials to calculate the electronic band structures for all cases. The spin-orbit coupling (SOC) is also included in the electronic band structure calculations. The harmonic force constants are calculated from density-functional perturbation theory (DFPT)\cite{fugallo2013ab}. The dielectric constant and Born effective charge are calculated to account for the long-range electrostatic interactions. Additionally, the Born–Huang rotational invariance correction and Huang conditions correction\cite{lin2022general} are applied to the phonon dispersion. The dynamical quadrupole tensors are calculated from linear response theory as implemented in ABINIT\cite{gonze2020abinit,romero2020abinit}. The electron−phonon coupling matrix elements are first calculated under coarse $\mathbf{k/q}$-point meshes and then interpolated to dense $\mathbf{k/q}$-point meshes based on maximally localized Wannier functions\cite{marzari2012maximally,ponce2016epw}. Additional computational details and convergence tests are presented in Sections II-IV of the Supplemental Material.

The optimized lattice structure from DFT and dynamical matrices from DFPT are employed as the initial input in the SSCHA calculation. A $6\times6\times1$ supercell is adopted to generate the stochastic ensembles, and the quantum nuclear effects are considered in the thermal amplitudes of random atomic configurations. To accelerate the evaluation of the energies and forces of the randomly generated atomic configurations, the neural equivariant interatomic potential (NequIP)\cite{batzner20223} is utilized and has a similar accuracy as the DFT calculations (see Section V of the Supplemental Material). The optimized dynamical matrices and lattice structure are obtained by minimizing the free energy. The QA effects are incorporated based on the "bubble" approximation\cite{monacelli2021stochastic}, and the fourth-order anharmonicity is shown to have little effect on the dynamical matrices. $5\times10^{4}$ configurations are used when calculating the renormalized dynamical matrices. More details about the SSCHA calculations are shown in Section VI of the Supplemental Material.

\begin{figure}[H]
    \includegraphics[width=1\columnwidth]{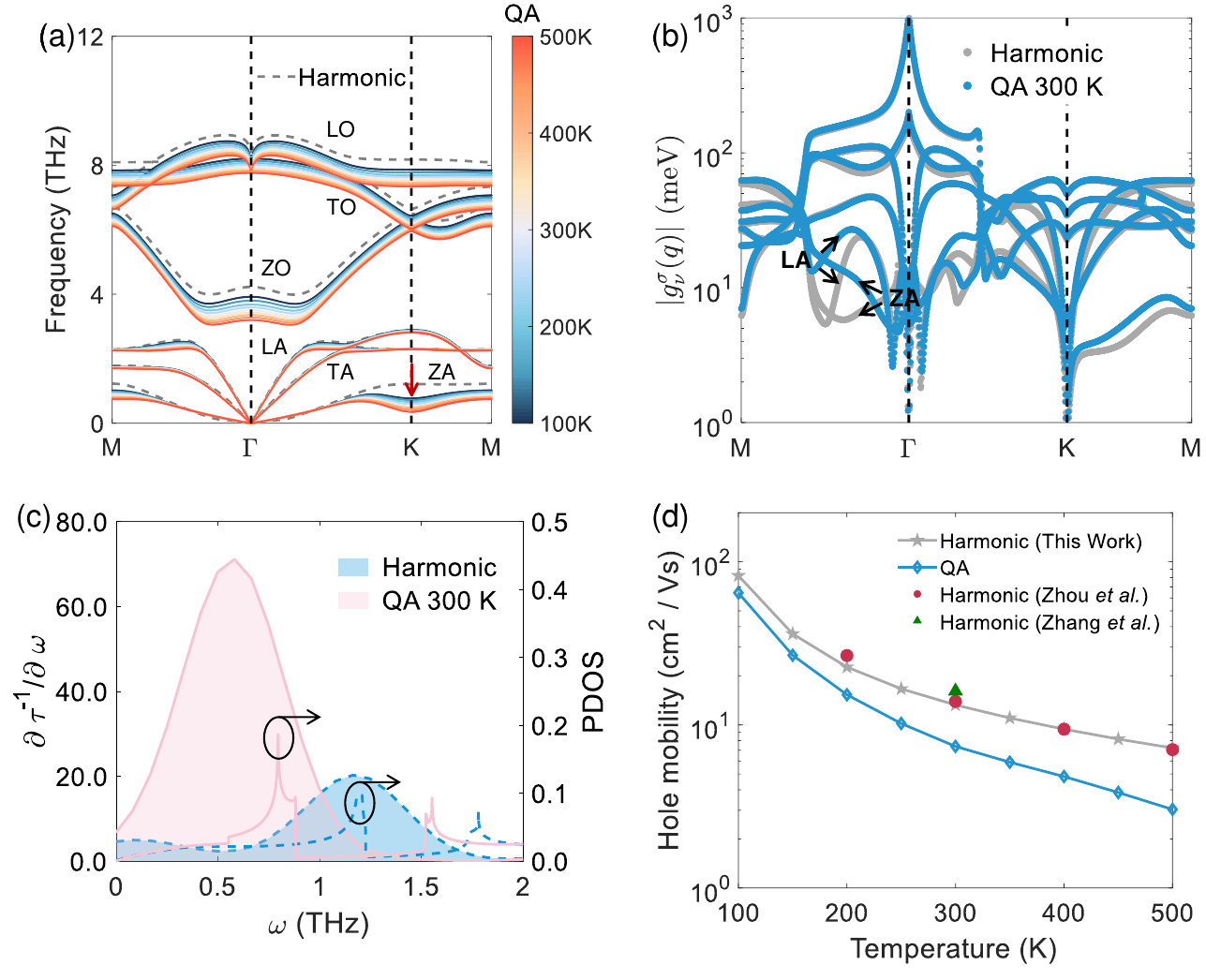}
    \caption {(a) Phonon dispersions of relaxed 2D-GaAs from the harmonic approximation (dashed lines) and the SSCHA incorporating QA effects. The phonon branches are labeled as out-of-plane acoustic (ZA), in-plane transverse acoustic (TA), in-plane longitudinal acoustic (LA), out-of-plane optical (ZO), in-plane transverse optical (TO), and in-plane longitudinal optical (LO). (b) The $\left|g_{v}^{\sigma}(\mathbf{q})\right|$ is calculated with harmonic/renormalized dynamical matrices. (c) Spectral decomposed hole scattering rates calculated with harmonic/renormalized dynamical matrices as a function of phonon energy calculated at 39 meV away from the valence band edges. The shaded regions indicate $\partial \tau^{-1} / \partial \omega$ (left axis), and the dashed lines represent the PDOS (right axis). (d) The $\mu_h$ as a function of temperature, compared with the data reported by Zhou \emph{et al}.\cite{zhou2024guidelines} (circle) and Zhang \emph{et al}.\cite{zhang2023two} (triangle).}
    \label{fig: Figure 1}
\end{figure}

\textit{Quantum anharmonicity effects on electron transport} - Fig. \ref{fig: Figure 1}(a) shows the phonon dispersion of relaxed 2D-GaAs from the harmonic approximation (DFPT) and the SSCHA incorporating the QA effects. It is found that there is no imaginary frequency in the harmonic phonon dispersion in the whole Brillouin zone. This is different from other strong anharmonic materials, like $\text{SrTiO}_3$\cite{zhou2018electron,ranalli2024electron} and $\text{KTaO}_3$\cite{ranalli2023temperature}, which have significant imaginary phonon frequencies. The QA effects would be tacitly ignored for 2D-GaAs at first glance. However, there is a strong renormalization on the harmonic dynamical matrices even at 100 K by incorporating the QA effects. The renormalization is especially strong on the ZA polarization and optical polarizations. In addition, the QA effects become more significant as the temperature increases. The ZA polarization is greatly softened at the K-point, as marked by the red arrow in Fig. \ref{fig: Figure 1} (a). Imaginary frequencies appear at the K-point when the temperature is higher than 700 K, as shown in Section VII of the Supplemental Material. This indicates that there is a phase change in 2D-GaAs in the temperature range of 600-700 K. Given that state-of-the-art monochromatic electron energy loss spectroscopy and scanning transmission electron microscopy technologies enable detecting phonon dispersions in 2D materials\cite{li2024observation}, the QA effects on lattice vibration of 2D-GaAs discussed here can be verified in future experiments.

 \begin{figure*}[hbpt]
    \centering
    \includegraphics[width=2.0\columnwidth]{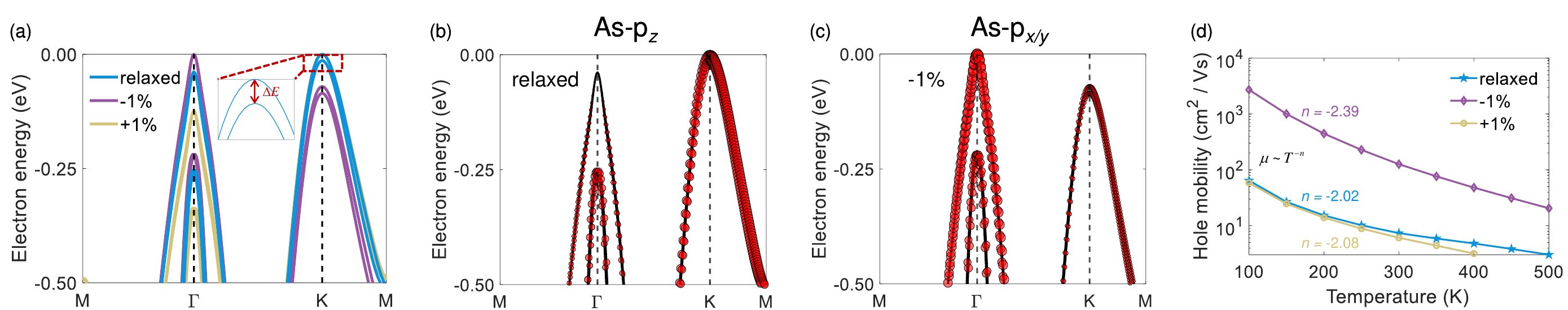}
    \caption {\justifying (a) Valence bands for 2D-GaAs with the relaxed geometry, the geometry with -1\% strain, and the geometry with +1\% strain. The inset shows the energy difference ($\Delta E$) of relaxed geometry induced by SOC. The electron energy is normalized to the VBM. Projection of (b) As-$p_{z}$ orbitals onto valence bands of the relaxed geometry and (c) As-$p_{x/y}$ orbitals onto valence bands of the -1\% strain geometry. (d) $\mu_h$ as a function of temperature for relaxed, -1\% strain, and +1\% strain geometries.}
    \label{fig: Figure 2}
\end{figure*}

To quantitatively investigate the QA effects on the electron-phonon interactions, the square root of the gauge-invariant trace ($\left|g_{v}^{\sigma}(\mathbf{q})\right|$) over the four highest valence bands at the $\Gamma$-point for dynamical matrices from both DFPT and SSCHA is shown in Fig. \ref{fig: Figure 1}(b). The coupling strength between electrons and ZA/LA polarizations is significantly enhanced due to the QA effects. 
On the other hand, the phonon density of state (PDOS) near the K-point is significantly increased due to the soft phonons induced by the QA effects, further resulting in the significant increase in spectral decomposed hole scattering rate, as shown in Fig. \ref{fig: Figure 1}(c). We examined the hole scattering rates regarding the ZA polarization (see Section VIII of the Supplemental Material). There is a significant increase in the hole scattering rates considering QA effects, especially for the hole states whose energies are 20 meV below the VBM. The $\mu_h$ calculated with the harmonic/renormalized dynamical matrices, as well as other existing theoretical data\cite{zhou2024guidelines,zhang2023two} calculated with the harmonic dynamical matrices, are shown in Fig. \ref{fig: Figure 1}(d). Our $\mu_h$ for the harmonic phonon case agrees quite well with data reported in Ref. \cite{zhou2024guidelines} over a wide temperature range. However, the $\mu_h$ calculated with the renormalized dynamical matrices is significantly smaller. The room-temperature $\mu_h$ is reduced by $\sim$44\% due to the QA effects. The reduction in $\mu_h$ can be even much larger at higher temperatures, as the lattice anharmonicity is significantly increased. Therefore, the QA effects are accounted for in the subsequent $\mu_h$ calculations of 2D-GaAs, unless otherwise specified.

\textit{Hole mobility of 2D-GaAs without/with biaxial strains} - 2D-GaAs exhibits one of the lowest $\mu_h$ among 2D semiconductors\cite{zhou2024guidelines,zhang2023two}. It should be emphasized that the $\mu_h$ of 2D-GaAs is further reduced to an extremely low value of 7.4 $\mathrm{cm\textsuperscript{2}/(Vs)}$ at room temperature when the QA effects are considered in our prediction. Therefore, it is highly desired to enhance the $\mu_h$ of 2D-GaAs. The fundamental reason for the low $\mu_h$ lies in the large electron scattering rate for electron states in the vicinity of the K-point. There are six-hole pockets in the first Brillouin zone due to the symmetry lattice, as shown by the Fermi surface in Fig. IX(a) in the Supplementary Material. The ZA phonon modes at the K-point satisfy the energy and momentum conservation for the interband electron-phonon scattering between electron states in the neighboring hole pockets. It should be noted that although the energy difference between the two valence bands near the VBM ($\Delta E$ in the inset of Fig. \ref{fig: Figure 2}(a)) is tiny, it is still much larger than the maximum phonon frequency of the ZA polarization. Therefore, the interband electron-phonon scattering between the two bands can be ignored.

Here, we propose a practical strategy to dramatically enhance $\mu_h$ through the valence band edge shift (VBES). Specifically, the sharp hole pocket at the $\Gamma$-point is shifted up and generates the new VBM. There is only one hole pocket at the $\Gamma$-point, as shown by the Fermi surface in Fig. IX(b) in the Supplementary Material. As a result, the electron-phonon scattering is dramatically decreased due to the full filtering of the original interband electron-phonon scattering channels that existed in the flat hole pocket. In practice, this VBES can be realized by strain engineering. As shown in Fig. \ref{fig: Figure 2}(a), a significant VBES appears when a 1\% biaxial compressive (-1\%) strain is applied. In contrast, there is no VBES under a 1\% tensile (+1\%) strain, as the hole pocket at the $\Gamma$-point is shifted down and the VBM at the K-point remains the same. Note that 2D-GaAs remains stable under strain. However, a phase change occurs at temperatures above 500 K for the -1\% strain and above 400 K for the +1\% strain, as shown in Section VII of the Supplemental Material. It should be noted that the valence band shift here is different from that in wurtzite GaN \cite{ponce2019route}, 4H-SiC\cite{sun2024giant}, diamond \cite{sun2025unlocking}, and silicon \cite{roisin2024phonon}. The valence band shift happens at the $\Gamma$-point in those 3D semiconductors. To uncover the underlying physics of the valence band shift in 2D GaAs, we further analyze the orbital-projected band structures shown in Figs. \ref{fig: Figure 2}(b) and \ref{fig: Figure 2}(c). For the relaxed geometry, the valence bands near the VBM are primarily composed of As-$p_{z}$ orbitals. Since the compressive strain is applied in the $xy$ plane, it significantly modulates the $p_{x/y}$ orbitals. Therefore, the energy of As-$p_{x/y}$ is correspondingly increased and becomes the dominant orbital component at the valence band edge, generating the new VBM at the $\Gamma$-point.

Fig. \ref{fig: Figure 2}(b) shows the $\mu_h$ of relaxed and strained 2D-GaAs. The room-temperature $\mu_h$ of the relaxed geometry is increased by $\sim$1600\% with a -1\% strain (from 7.4 to 126.4 $\mathrm{cm\textsuperscript{2}/(Vs)}$), making $\mu_h$ the same magnitude as $\mu_e$. In contrast, the $\mu_h$ is decreased by $\sim$17.6\% with a +1\% strain, which is related to the decrease of the contributions from holes with small effective mass. Typically, $\mu_h$ exhibits an exponential decrease with temperature ($\mu \sim T^{-n}$), as phonons are significantly activated at high temperatures. The temperature dependence of $\mu_h$ becomes more pronounced, with the exponent \textit{n} changing from -2.02 to -2.39 when a -1\% strain is applied. Although the relative increase in $\mu_h$ with compressive strain is decreased at higher temperatures, it is still considerable and achieves $\sim$578\% at 500 K. In addition, we examined the effect of strain on the conduction bands of 2D-GaAs, as shown in Section X of the Supplemental Material. There is little change in the conduction band edge with strain, and it can be deduced that the impact of strain on the $\mu_e$ of 2D-GaAs is negligible.

\begin{figure}[H]
    \includegraphics[width=1.0\columnwidth]{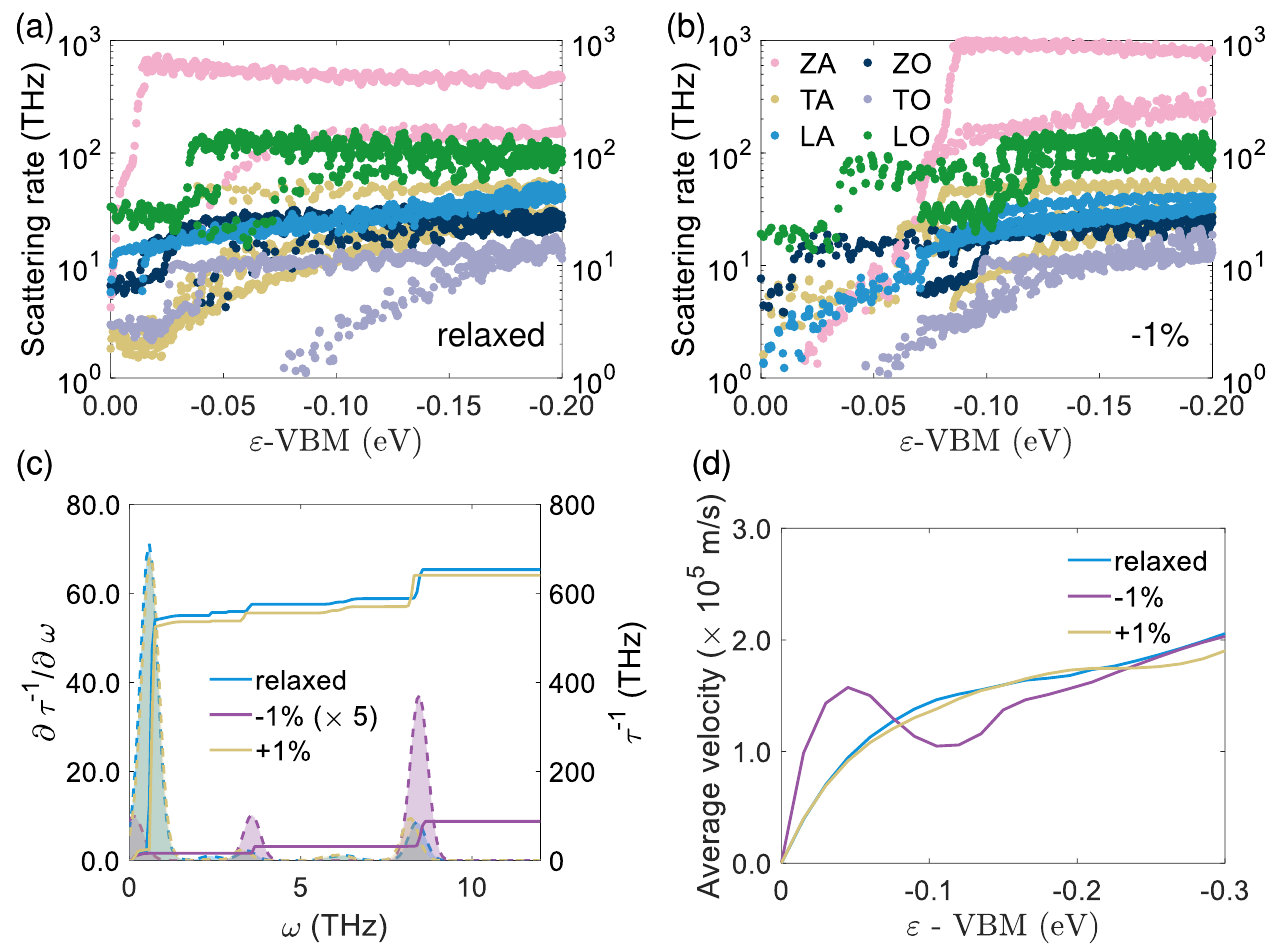}
    \caption{\justifying Mode-resolved room-temperature hole scattering rates with respect to the energy below the VBM of (a) relaxed and (b) -1\% strain 2D-GaAs. (c) Spectral decomposed hole scattering rates for 2D-GaAs without/with strains as a function of phonon energy calculated at 39 meV away from the valence band edges. The peaks indicate $\partial \tau^{-1} / \partial \omega$ (left axis), and the solid lines represent the cumulative integral $\tau^{-1}$ (right axis). Note that the spectral decomposed hole scattering rates under -1\% strain are presented with a fivefold magnification to highlight their features. (d) The average electron group velocities for 2D-GaAs without/with strains. The average electron group velocity is calculated according to the equation $\bar{v}^{\alpha}(\varepsilon)=\sum_{n\mathbf{k}}\left|v^{\alpha}_{n\mathbf{k}}\right| \delta\left(\varepsilon-\varepsilon_{n\mathbf{k}}\right) / \sum_{n\mathbf{k}} \delta\left(\varepsilon-\varepsilon_{n\mathbf{k}}\right)$.}
    \label{fig: Figure 3}
\end{figure}

To reveal the microscopic mechanisms regarding the dramatic enhancement in $\mu_h$, the hole scattering rates for relaxed and -1\% strain geometries are shown in Figs. \ref{fig: Figure 3}(a-b), respectively. It is found that scattering rates from the ZA polarization are severely suppressed under -1\% strain. It should be mentioned that LO polarization dominates the hole scattering in -1\% strain geometry due to the large reduction of hole scattering from ZA polarization. This can be further validated by the spectral decomposed angular-averaged scattering rate at a phonon energy of 3\textit{$k_b$}\textit{T}/2 (39 meV at 300 K) away from the VBM, as shown in Fig. \ref{fig: Figure 3}(c). The most hole scatterings are attributed to acoustic phonons for both relaxed and +1\% strain geometries, while high-frequency optical phonons are dominant for the -1\% strain geometry. 

The changes in the profile of valence band edges can also have significant effects on the electron group velocity, which is shown in Fig. \ref{fig: Figure 3}(d). The average electron group velocity for electron states near the VBM under a -1\% strain is significantly higher than that of the relaxed geometry. This is attributed to the sharp valence band edge of the compressive geometry. It should be noted that there is a sudden decrease in the group velocity for electron states $\sim$0.1 eV below the VBM, which is ascribed to the relatively flat hole pocket at the K-point. This decrease does not impede electron transport significantly as those electron states have little contribution to $\mu_h$. There is little change in the electron group velocity of the +1\% strain geometry. This is consistent with the unchanged VBM at the K-point with/without tensile strain, as shown in Fig. \ref{fig: Figure 2}(a). Therefore, the dramatic increase in $\mu_h$ under compressive strain is attributed to the synergistic enhancement of electron relaxation time and group velocity, but the former is the main reason.

Finally, we explore the practical approaches to realizing the strain engineering in 2D semiconductors. Experimentally, strain is primarily induced through interactions with substrates, such as epitaxial growth, thermal expansion mismatch, and substrate surface topography modification\cite{qi2023recent}. However, the weak van der Waals interactions and significant slippage between the 2D semiconductor and substrate limit the effective transfer and control of strain\cite{deng2018strain}. Recently, the emergence of polymer encapsulation methods\cite{li2020efficient} has addressed these challenges, providing a more reliable solution for strain engineering. It should be noted that there is a critical uniaxial compressive strain for the reversal of valence band order in 3D wurtzite GaN\cite{ponce2019route}, 4H-SiC\cite{sun2024giant}, and rocksalt ScN\cite{rudra2023reversal}, which is still challenging in experimental realization. In contrast, the VBES in 2D-GaAs operates without the need for such a critical compressive strain. It can happen with even a small 0.1\% compressive strain, which indicates that the enhancement in $\mu_h$ of 2D-GaAs can be practically achieved in experiments. Furthermore, we confirm that the VBES phenomenon also occurs in other 2D III-V semiconductors by checking the biaxial strain induced variations in the valence band edges of 2D-InP and 2D-GaP (see Section XI of the Supplemental Material). However, we found that the moderate biaxial compressive strain is inefficient to enhance the $\mu_{h}$ of other groups of 2D semiconductors, like $\mathrm{MoS_{2}}$ and phosphorene (see Sections XII and XIII of the Supplemental Material).

In summary, we conducted a comprehensive first-principles investigation on the electron-phonon interactions and hole transport in 2D III-V semiconductors. The QA effects are incorporated via the stochastic self-consistent harmonic approximation assisted by the machine learning potential. It is found that the QA effects significantly renormalize the dynamical matrices from the harmonic approximation. The hole mobility of 2D-GaAs is reduced by a $\sim$44\%, reaching an extremely low value of 7.4 $\mathrm{cm\textsuperscript{2}/(Vs)}$, as a result of enhanced scattering from the ZA polarization caused by the renormalized dynamical matrices. We proposed the valence band edge shift strategy to dramatically increase the hole mobility, which can be realized by a slight biaxial compressive strain. The scattering channels corresponding to original interband electron–phonon scattering channels in the flat hole pocket are fully filtered as the strain is applied, leading to a significant reduction in the electron–phonon scattering rate. In addition, the electron group velocity is considerably promoted due to the decrease of the hole effective mass. This work provides profound insights into the hole transport in 2D semiconductors, which can facilitate their scalable applications in nanoelectronics. 

S.L. was supported by the National Natural Science Foundation of China (Grant No. 12304039) and the Shanghai Municipal Natural Science Foundation (Grant No. 22YF1400100). X.L. was supported by the Shanghai Municipal Natural Science Foundation (Grant No. 21TS1401500) and the National Natural Science Foundation of China (Grants Nos. 52150610495 and 12374027). J.S. was supported by the Fundamental Research Funds for the Central Universities (Grant No. CUSF-DH-T-2024061). The computational resources utilized in this research were provided by National Supercomputing Center in Shenzhen.

\bibliography{bibliography}
\end{document}